\documentclass[12pt]{article}

\usepackage{amsfonts,latexsym,eucal,color,graphicx}
\usepackage{epsfig,amssymb,amsmath,mathrsfs,tabularx,bm,cite}

\newenvironment{minilinespace}{\baselineskip = 8mm}{}

\begin{document}

\begin{titlepage}

\begin{flushright}
{ \small
	arXiv:0709.1028 [hep-th]\\
	WU-AP/270/07
}
\end{flushright}

\vspace{1cm}

\begin{minilinespace}
\begin{center}
	{
		\Large
		{\bf
			Analytic evidence for
			the Gubser-Mitra conjecture
		}
	}
\end{center}
\end{minilinespace}
\vspace{1cm}

\begin{center}
Umpei Miyamoto\\
\vspace{.5cm}
{\small \textit{
Department of Physics,
Waseda University, Okubo 3-4-1, Tokyo 169-8555, Japan
}}
\\
\vspace*{1.0cm}

{\small
{\tt{
\noindent
umpei@gravity.phys.waseda.ac.jp
\\
}}
}
\end{center}

\vspace*{1.0cm}



\begin{abstract}
A simple master equation for the static perturbation of charged black strings is derived while employing the gauge proposed by Kol. As the charge is varied it is found that the potential in the master equation for the perturbations becomes positive exactly when the specific heat turns positive thus forbidding a bound state and the onset of the Gregory-Laflamme instability. It can safely be said that this is the first analytic and explicit evidence for the Gubser-Mitra conjecture, correlating the classical and thermodynamic instabilities of black branes. Possible generalizations of the analysis are also discussed. 
\end{abstract}

\end{titlepage}

\section{Introduction}

It has been known that black objects with an extended event horizon, such as black branes and strings, suffer from a classical instability, called the Gregory-Laflamme (GL) instability~\cite{Gregory:1993vy,Gregory:1994bj}. The ultimate fate of this instability remains a matter of investigation in spite of extensive studies (see Refs.~\cite{Kol:2004ww,Harmark:2007md} for reviews and related works). One of the interesting aspects of this instability is the correlation with the local thermodynamic instability of background spacetimes.
This correlation of instabilities is known as the Gubser-Mitra (GM) conjecture \cite{Gubser:2000ec,Gubser:2000mm} (or the correlated stability conjecture~\cite{Gubser:2004dr}): the GL instability for black objects with a non-compact translational symmetry occurs if and only if the background black objects are locally thermodynamically unstable. Supports based on a general semiclassical argument~\cite{Reall:2001ag,Harmark:2005jk} and a lot of evidences have been known for this conjecture~\cite{Gubser:2000ec,Gubser:2000mm,Hirayama:2002hn,Gubser:2004dr,Miyamoto:2006nd}.~\footnote{It should be mentioned that possible counterexamples and refinement of the conjecture have been proposed~\cite{Friess:2005zp,Marolf:2004fya}.
In this article, however, we only consider the original version, presented above.
}
However, mainly due to the difficulty to solve the complicated perturbation equations of black branes analytically, any conclusive proof of the conjecture has not been reported, as far as the author knows.
Since the GM conjecture plays crucial roles in the stability argument of black objects in string theory as well as in the holographically dual gauge theories~\cite{Aharony:2004ig,Buchel:2005nt}, it is important to prove the conjecture or have analytic evidence for it. 

Now we turn to the complexity/simplicity of the perturbation equations of black branes. Much effort has been devoted to simplifying the perturbation equations and to identifying analytically the GL modes. Among them, the expression of perturbation equation obtained by Kol \cite{Kol:2006ga} would be simplest. See \cite{Kol:2006ga} and references therein for the comparison between various gauges adopted in the literature. His gauge choice was optimized by the action formalism in which gauge fixing is postponed as much as possible and the action is transformed into a canonical form making use of its own invariance~\cite{Kol:2006ga}. The resultant expression of perturbation equation is so simple that it allows for analytic investigations or accurate numerical ones. For example, an approximation formula of the dimensional dependence of GL marginal mode, which is sufficiently accurate in all dimensions, was obtained in~\cite{Asnin:2007rw},
while solving the perturbation equation for general dimensions is difficult even in such a gauge.

In this letter, we adopt the gauge mentioned above for the static perturbation of (non-dilatonic) magnetically charged black strings to obtain a single master equation. We show that if such a master equation is written in Schr\"{o}dinger form, the potential exhibits a significant feature:~the potential becomes positive definite to forbid the critical mode of GL instability for locally thermodynamically stable black strings. That is, one sees an explicit realization of the GM conjecture in this system.

\section{ Magnetic black strings }
\label{sec:setup}

We consider the following $ ( d+1 ) $-dimensional action ($ d \geq 4 $),
\begin{eqnarray}
	I
	=
	\frac{1}{16\pi G} \int d^{d+1}x \sqrt{-g}
		\left[
			R-\frac{1}{2(d-2)!} F_{d-2}^{2}
		\right]\ ,
	\label{eq:action}
\end{eqnarray}
where $ F_{d-2} $ is a $(d-2) $-form gauge field. Varying action (\ref{eq:action}), we have the following EOMs,
\begin{eqnarray}
	&&
	R_{\mu\nu} 
	 =
	\frac{1}{2(d-3)!} { F }_{ \mu }^{ \;\;\mu_2 \ldots \mu_{d-2} }
		{ F }_{ \nu \mu_2 \ldots \mu_{d-2} } - \frac{ d-3 }{ 2(d-1)! }
		g_{ \mu\nu } F^{2}\ ,
	\label{eq:ein}
	\\
	&&	
	\nabla_{\mu} F^{ \mu\mu_{2} \ldots \mu_{d-2} } = 0\ .
	\label{eq:maxwell}
\end{eqnarray}
The form field must satisfy Bianchi identity, $  \mathrm{ d } F_{ d-2 }  = 0 $, in addition to Eq.~(\ref{eq:maxwell}). By a dimensional reduction method, a black string solution in this system can be obtained from a black hole solution in a $d$-dimensional dilatonic system~\cite{Horowitz:1991cd}. The explicit form of the solution is
\begin{eqnarray}
	&&
	ds^2
	=
	- f_+ dt^2
	+ \frac{ dr^2 }{ f_+ f_- } 
	+ f_- dz^2
	+ r^2 d\Omega_{ d-2 }^2\ ,
	\;\;\;
	f_{\pm}(r) = 1 -\left( \frac{r_{\pm}}{r} \right)^{d-3}\ ,
	\nonumber
	\\
	&&
	F = \tilde{Q}  \varepsilon_{d-2}\ ,
	\;\;\;
	\tilde{Q} = \pm \sqrt{ (d-1)(d-3) } ( r_+ r_- )^{ (d-3)/2 }\ ,
	\label{eq:BG}
\end{eqnarray}
where $ d\Omega_{d-2}^2 $ and $ \varepsilon_{ d-2 } $ are the line and volume elements of a unit $(d-2)$-sphere, respectively. In addition, $r=r_+$ and $r=r_-$ correspond to an event horizon and inner horizon, respectively.
To make physical quantities finite, let the spacetime be compact in $z$-direction with length $L$. Then, mass, temperature and magnetic charge are calculated as
\begin{eqnarray}
	&&
	M 
 	= 
	\frac{ \Omega_{d-2} L r_+^{d-3} }{ 16 \pi G } 
  			\left(
 				d-2 + q^{d-3}
  			\right)\ ,
  	\;\;\;\;
	T
	=
	\frac{ d-3 }{ 4\pi r } \sqrt{ f_- }\Big|_{r=r_+}\ ,
	\nonumber
	\\
	&&
	\hspace{3cm}
	Q
	=
	\frac{  \Omega_{d-2} L r_+^{d-3} }{ 16 \pi G } (d-1) q ^{ (d-3)/2 }\ ,
 \label{eq:MTQ}
\end{eqnarray}
where $\Omega_{d-2}$ is the surface area of unit $(d-2)$-sphere. $q$ is a charge (or extremal) parameter, defined by
\begin{eqnarray}
	q \equiv \frac{r_-}{r_+}\ ,
	\;\;\;\;\;
	( 0 \leq q < 1 )\ .
\end{eqnarray}
Note that the magnetic charge in Eq.~(\ref{eq:MTQ}) is normalized so that $Q \to M$ in the extremal limit, $q\to 1$.
We can calculate the specific heat for the above black string,
\begin{eqnarray}
 &&
 	C_{Q} 
 	=
 	\left(
 		\frac{ \partial M }{ \partial T }
 	\right)_{Q}
 	=
 	\frac{ \Omega_{d-2} L r_+^{d-2} }{ 4 G }
 	\frac{ \sqrt{ 1-q^{d-3} } [(d-2) - q^{d-3}] }{ -1 + (d-2)q^{d-3}}\ .
 \label{eq:SH}
\end{eqnarray}
We can see from Eq.~(\ref{eq:SH}) that there is a critical value of the charge parameter, we denote it by $q_{\mathrm{GM}}$, above which the specific heat becomes positive,
\begin{eqnarray}
    q_{\mathrm{GM}} \equiv \frac{1}{(d-2)^{1/(d-3)}}\ .
 \label{eq:qc}
\end{eqnarray}
The GM conjecture asserts that the GL instability does not exist for $ q > q_{\mathrm{GM}} $.
We note that this criterion of thermodynamic stability corresponds to the one in a canonical ensemble. If a magnetic charge is allowed to be re-distributed in $z$-direction, which is not the case in the present spacetime, we have to take into account also the positivity of isothermal permittivity, corresponding to working in a grandcanonical ensemble~\cite{Harmark:2007md}. 

\section{ Static perturbation }
\label{sec:pert}

\subsection{ Gauge choice }
\label{sec:gauge}

According to \cite{Kol:2006ga}, we consider the following ``maximally general ansatz'',
\begin{eqnarray}
	ds^2
	=
	- f_+ e^{ 2 a(r,z) } dt^2
	+ \frac{ e^{ 2b(r,z) } }{ f_+ f_- } dr^2
	+ f_- e^{ 2 \beta }
		\left[
			dz - \alpha(r,z) dr
		\right]^2
	+ r^2 e^{ 2c(r,z) } d\Omega_{ d-2 }^2\ .
	\label{eq:metric2}
\end{eqnarray}
We regard all metric functions introduced in Eq.~(\ref{eq:metric2}), $(a,b,c,\alpha,\beta)$, as small perturbations. Since metric (\ref{eq:metric2}) has extra degrees of freedom to describe a non-uniform black string, one can impose gauge conditions to restrict the perturbations to be physically relevant. 
For example, a conformal-type gauge in ($r,z$) plane, defined by $ b=\beta $ and $\alpha=0$, was adopted in Ref.~\cite{Gubser:2001ac}. In this gauge, $b$ can be written in terms of $a$ and $c$, and we have coupled equations for $a$ and $c$ (after expanding each function by suitable harmonics in $z$-direction). It is known that this gauge equally goes well for the charged case~\cite{Miyamoto:2006nd}. On the other hand, the ``optimal gauge'' proposed in Ref.~\cite{Kol:2006ga} for the static perturbations of neutral black strings ($q=0$ in the above solution) is to set $ a = \beta = 0 $ with $ b $, $c$ and $\partial_z \alpha$ taken as variables. In this gauge, one can show that both $ b $ and $ \partial_z \alpha $ are ``non-dynamical variables'', whose derivatives do not enter into EOMs and can be written in terms of master variable $c$ (see \cite{Kol:2006ux} for a general description of such gauge optimization and decoupling procedure).
Also in the present charged case ($ q\neq 0$), in which the Einstein equation has a source term of the gauge field, we can take the same gauge to obtain a master equation for $c$. It is not so trivial that this gauge goes well even in the charged case. The point is that under ansatz (\ref{eq:metric2}), the gauge field $ F = \tilde{Q} \epsilon_{d-2} $ in Eq.~(\ref{eq:BG}) remains to be a solution, i.e., EOM (\ref{eq:maxwell}) and the Bianchi identity are satisfied in this form. As a result, the right hand side of Einstein equation (\ref{eq:ein}) contains only $c$, e.g., $ F^2 \propto \tilde{Q}^2/(re^c)^{2(d-2)} $.
In other words, the perturbation of form field can be written in terms of the perturbation of the metric.
Thus, the gauge field neither introduces any unknown function nor changes the structure of the Einstein equations.

Now, let us normalize the coordinates by the horizon radius,
\begin{eqnarray}
	y \equiv \frac{r}{r_+}\ ,\;\;\;\;\;
	x \equiv \frac{z}{r_+}\ .
\end{eqnarray}
Then, we take the gauge of $ a=\beta=0$ and move to a Fourier space by
\begin{eqnarray}
	&&
	X (y,x)
	=
	X_1 (y) \cos ( k x )\ ,
	\;\;\;\;\;\;
	( X = b, c)\ ,
	\nonumber
	\\
	&&
	\alpha (y,x)
	=
	\alpha_1 ( y ) \partial_x \cos ( k x)\ ,
	\label{eq:expand2}
\end{eqnarray}
where $ k $ is GL critical wavenumber to be determined later.
Substituting (\ref{eq:expand2}) into Einstein equation (\ref{eq:ein}), we have three independent equations, which do not contain the derivatives of $b_1$ and $\alpha_1$.
Using two of these three equations,  both $b_1$ and $\alpha_1$ can be written in terms of $c_1$ (and its first derivative). Finally, the master equation for $c_1$ is obtained, which we will investigate in the rest part of this section.

\subsection{ Non-existence of critical mode for $q>q_{\mathrm{GM}}$ }
\label{sec:gauge}

The master equation for $c_1$ in the potential form is given by
\begin{eqnarray}
	&&
	\left[
		- e^{ -{\mathcal{A}} -{\mathcal{B}} }
		\partial_y  ( e^{  {\mathcal{A}} -{\mathcal{B}}} \partial_y )
		+
		V(y)
	\right] c_1
	=
	-k^2 c_1\ ,
	\label{eq:master}
\end{eqnarray}
where
\begin{eqnarray}
	&& 
	e^{  2{\mathcal{A}}} 
	=
	\frac{ y^{2d} f_+ {f_+'}^4}{ [2(d-2) f_{+} + yf_+' ]^4}\ ,
	\;\;\;\;\;\;\;
	e^{  - 2{\mathcal{B}}}
	=
	f_+ f_-^2\ ,
	\label{eq:op}
	\\
	&&
	V(y)
	=
	\frac{ 2( d-3 ) f_{-} 
		\left\{ 3-d
			-[
				d^2-4d+5
				-(d-1)(d-2)f_-
			] f_+
		\right\}
			}
			{
				y^2 [ d-3 + ( d-1 ) f_{+} ]
			}
	\nonumber
	\\
	&&
	\hspace{8cm}
	-
	\frac{
			2 ( d-2 ) \tilde{Q}^2 { f_- } 
		}
		{
			( d-1 ) y^{ 2( d-2 ) } ( f_{+} -1 ) 
		}\ .
	\label{eq:pote}
\end{eqnarray}
The second term in potential (\ref{eq:pote}) comes from the source term of the Einstein equation. Substituting the explicit form of $ f_\pm $ and $ \tilde{Q} $, Eq.~(\ref{eq:pote}) can be written in a simpler form,
\begin{eqnarray}
	V(y)
	=
	\frac{ 
		2(d-3)^2
		\left[
			-1+(d-2)q^{d-3}
		\right] 
	(y^{d-3}-q^{d-3})
	}
	{
		\left[ 2(d-2)y^{d-3} - (d-1) \right] y^{d-1}
	}\ .
	\label{eq:pote2}
\end{eqnarray}

\begin{figure}[t]
	\begin{center}
	\begin{tabular}{ cc }
	\includegraphics[width=6.5cm]{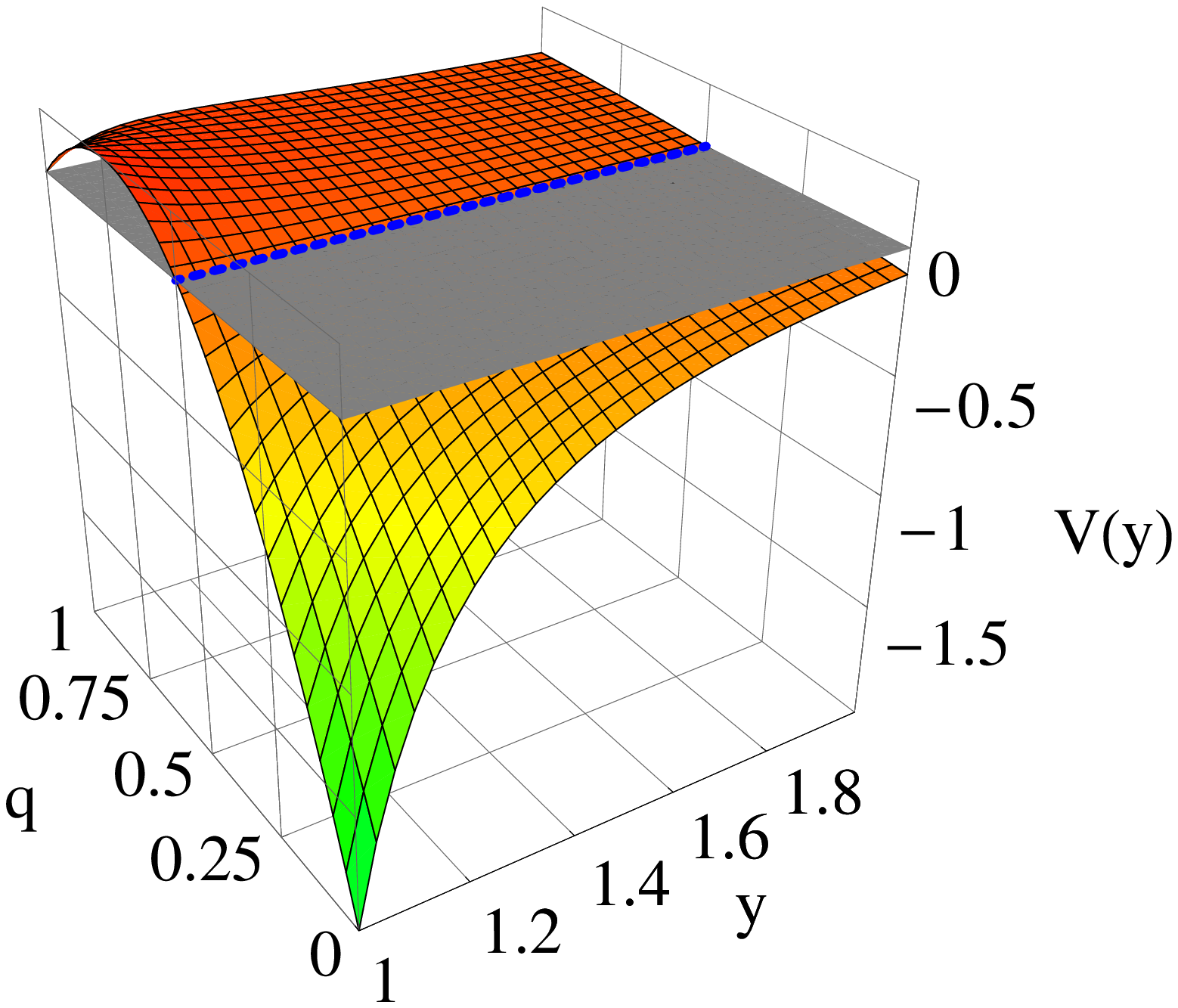} &
	\includegraphics[width=8cm]{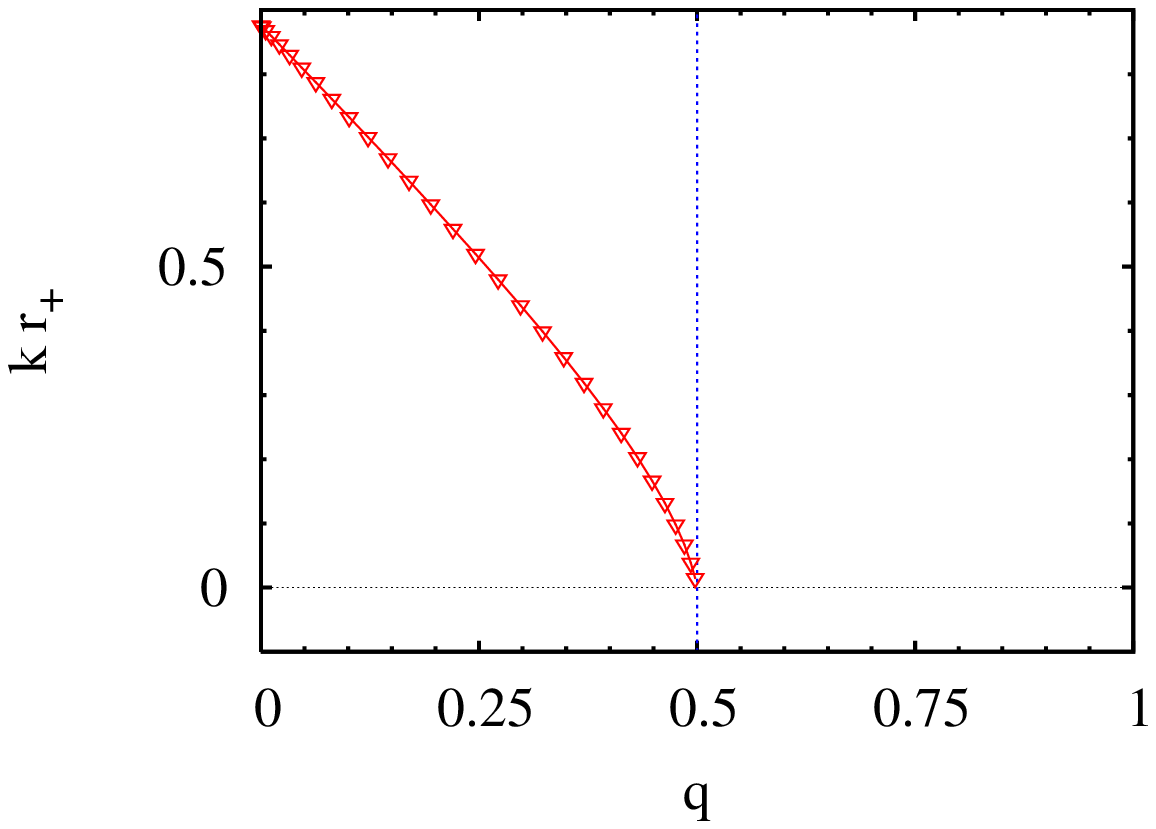} \\
	(a) & (b)
	\end{tabular}
		\caption{ \small
(a) The charge dependence of potential~(\ref{eq:pote2}) for $d=4$. The qualitative feature is the same in the other dimensions ($d \geq 5$). The event horizon is located at $y=1$ and $ q = 1 $ corresponds to the extremal limit. The background black string is thermodynamically stable for $ q > q_{ \mathrm{GM} } = 1/(d-2)^{1/(d-3)} $, ($q_{\mathrm{GM}}=1/2$ for $d=4$). The potential (colored surface with mesh) intersects with the zero plane (gray plane) on $ q = q_{\mathrm{GM}} $ (blue dotted line) and is positive definite for $ q > q_{\mathrm{GM}} $.
(b) Numerical plot of the charge dependence of GL critical wavenumber (red triangles) for $d=4$. The curve is just to guide the eye. The critical wavenumber decreases monotonically as the charge increases and vanishes at $ q = q_{\mathrm{GM}} $ (blue dotted line). The non-existence of $ k \in \Re $ for $ q > q_{\mathrm{GM}} $ is analytically show by Eq.~(\ref{eq:vanish}).
\label{fg:pote} }
	\end{center}
\end{figure}

From Eq.~(\ref{eq:pote2}), one can see that the potential is regular and finite everywhere between the horizon and infinity, $ 1 \leq y < +\infty$. The behavior of potential in the asymptotic region is $ V(y) \sim y^{-(d-1)} $.~\footnote{The neutral limit ($q\to 0$) of potential (\ref{eq:pote}) or (\ref{eq:pote2}) does not coincide with the potential for the static perturbations of a neutral black string obtained in Ref.~\cite{Kol:2006ga}. This stems from the fact that the potential in \cite{Kol:2006ga} is not for $c$ itself but for a linear combination of $a $ and $c$. }
The most interesting property of the potential which can be seen from Eq.~(\ref{eq:pote2}) is
\begin{eqnarray}
	q \lesseqqgtr q_{\mathrm{GM}}
	\;\;
	\Longleftrightarrow
	\;\;
	V(y) \lesseqqgtr 0\ .
\end{eqnarray}
This charge dependence of the potential is visualized in Fig.~\ref{fg:pote} (a).
Using this significant property, one can show the non-existence of GL critical mode for $ q  > q_{\mathrm{GM}}$ as follows. Multiplying Eq.~(\ref{eq:master}) by $ e^{ \mathcal{A} + \mathcal{B} } c_1 $ and integrating it by parts, we have 
\begin{eqnarray}
	\int_{1}^{\infty}
		e^{ \mathcal{A-B} }
	\left[
		c_1^{\prime 2} + e^{2\mathcal{B}} (V + k^2) c_1^2
	\right] dy 
	-
	\left[
		e^{ \mathcal{A-B} } c_1 c_1^\prime
	\right]_{ y=1 }^{ y=\infty }
	=
	0\ .
	\label{eq:vanish}
\end{eqnarray}
The boundary term in Eq.~(\ref{eq:vanish}) vanishes. Then, it is obvious that the critical wavenumber $ k \in \Re $ does not exist for $ V(y) > 0 $. This is the analytic evidence that the GL instability cannot exist for the thermodynamically stable black strings in this system.

Now, the proof of the GM conjecture should be completed by showing the existence of the critical mode for $ 0 \leq q < q_{\mathrm{GM}} $.
However, it seems not so easy to solve Eq.~(\ref{eq:master}) analytically even in the simplest case $d=4$, and we have to resort to numerical computations. Therefore, we integrate Eq.~(\ref{eq:master}) numerically for $ 0 \leq q < q_{\mathrm{GM}} $ with regularity conditions at the horizon and infinity. In Fig.~\ref{fg:pote} (b), we show the charge dependence of the critical wavenumber $k$. We can see that $k$ indeed decreases monotonically as $q$ increases and vanishes almost exactly at $q=q_{\mathrm{GM}}$. See Refs.~\cite{Gubser:2002yi,Miyamoto:2006nd} for the observation that the vanishing of the critical wavenumber  around $ q_{\mathrm{GM}} $ exhibits a power-low behavior with universal critical exponent $1/2$, i.e., $ k \propto | Q_{\mathrm{GM}}- Q |^{1/2}$, where $Q_{\mathrm{GM}}$ is the physical charge corresponding to $q_{\mathrm{GM}}$.

\section{ Discussion }
\label{sec:conclusion}

We have shown that magnetically charged black string (\ref{eq:BG}) serves as an analytic evidence for the Gubser-Mitra conjecture. To prove the conjecture in a complete form, the partial proof given in Sec.~\ref{sec:gauge} should be followed by an analytic identification of the Gregory-Laflamme critical mode ----- alternatively, just by showing the existence of the mode analytically ----- for the thermodynamically unstable strings ($ 0 \leq q < q_{\mathrm{GM}} $). The extension to dynamical perturbation might be important, although we have only considered the marginally stable critical mode.

Here, we mention that perturbation (\ref{eq:expand2}) can be generalized to higher-order ones, taking into account the non-linear backreactions as developed in \cite{Gubser:2001ac}. In fact, we did confirm a single master equation to be derived at each order of perturbation for the charged strings considered in this paper. Technically speaking, this fact reduces the number of parameters to be determined numerically to one for each mode and saves the computation time considerably.~\footnote{Precisely speaking, such master equations were derived only for Kaluza-Klein ($k\neq 0$) modes. For homogeneous ($k = 0$) modes, which inevitably appear in the non-linear perturbations, we were not able to obtain the master equation, which was done for the neutral black string in \cite{Kol:2006ga}.
Instead, we have a set of coupled equations for $c$ and $\beta$ in the gauge of $ b = \alpha = 0 $ ($a$ is a non-dynamical variable to be written in terms of $c$ and $\beta$). However, we have only one shooting parameter even for such coupled equations.} We will discuss such non-linear perturbations elsewhere, by which we could confirm the critical dimension \cite{Sorkin:2004qq,Kudoh:2005hf} and critical charges \cite{Miyamoto:2006nd} with more accurate numerics/much higher-order perturbations.

Besides the black objects with the translational symmetries, the correlation between the instabilities is relevant to a black string in Anti-de Sitter spacetime \cite{Gregory:2000gf,Hirayama:2001bi}, having no translational symmetry, as well as to de Sitter spacetime \cite{Kinoshita:2007ci}. It would be interesting to generalize the gauge optimization in~\cite{Kol:2006ga} to these spacetimes to understand/refine the GM conjecture.

\section*{Acknowledgments} 
The author would like to thank H.~Kudoh for collaborating at an early stage of this work, and V.~Asnin, B.~Kol and N.~A.~Obers for useful discussions and comments.
The author thanks also to the organizers of the workshop ``Einstein's Gravity in Higher Dimensions'' (The Hebrew University of Jerusalem, Feb.~2007), where
this work was finalized by useful discussions with participants.
This work is supported by a grant from the 21st Century COE Program (Holistic Research and Education Center for Physics of Self-Organizing Systems) at Waseda University.



\bibliographystyle{JHEP}

\providecommand{\href}[2]{#2}\begingroup\raggedright\endgroup


\end{document}